\documentclass{article}

\usepackage[final, nonatbib]{nips_2018}

\usepackage[utf8]{inputenc} 
\usepackage[T1]{fontenc}    
\usepackage{hyperref}       
\usepackage{url}            
\usepackage{booktabs}       
\usepackage{amsfonts}       
\usepackage{nicefrac}       
\usepackage{microtype}      
\usepackage{graphicx}
\usepackage{amsmath}
\usepackage{amssymb}

\usepackage{todonotes}

\title{Structure-Based Networks for Drug Validation}

\author{
  C\u{a}t\u{a}lina Cangea$^1$, Arturas Grauslys$^2$, Pietro Li\`{o}$^1$ and Francesco Falciani$^2$\\
  $^1$University of Cambridge \quad $^2$University of Liverpool}

\begin{document}

\maketitle

\begin{abstract}
Classifying chemicals according to putative modes of action (MOAs) is of paramount importance in the context of \emph{risk assessment}. However, current methods are only able to handle a very small proportion of the existing chemicals. We address this issue by proposing an \emph{integrative deep learning architecture} that learns a joint representation from molecular structures of drugs and their effects on human cells. Our choice of architecture is motivated by the significant influence of a drug's chemical structure on its MOA. We improve on the strong ability of a unimodal architecture (F1 score of 0.803) to classify drugs by their \emph{toxic MOAs} (Verhaar scheme) through adding another learning stream that processes transcriptional responses of human cells affected by drugs. Our integrative model achieves an even higher classification performance on the LINCS L1000 dataset---the error is reduced by 4.6\%. We believe that our method can be used to extend the current Verhaar scheme and constitute a basis for fast drug validation and risk assessment.
\end{abstract}

\section{Introduction}

Industrial chemistry is nowadays heavily centered around two topics of increasing concern: risk assessment and drug regulation. Tens of thousands of new chemicals are being synthesised every year---this calls for a faster validation process, in order to advance possible candidates to the next drug development stage or ascertain toxicity levels before releasing chemicals into the environment.

One well-established risk assessment method is the Verhaar scheme~\cite{verhaar1992classifying}, which assigns a chemical one of four possible toxic modes of action: \emph{inert}, \emph{less inert}, \emph{reactive}, \emph{specifically acting}. This method is entirely based on chemistry principles and determines the class of a compound using a sequence of structural triggers that attempt to place a compound in one of the four classes~\cite{enoch2008classification}. However, the scheme can only be applied to a very small percentage of the existing chemical space, as the triggers do not have any effect on most compounds, which makes it impossible to determine their toxicity.

We propose an integrative method of learning from the \emph{transcriptional responses of human cells exposed to chemicals} and \emph{structural representations of the chemicals}. Our deep learning architecture achieves a high Verhaar classification performance (F1 score of 0.812) on chemicals used in the LINCS L1000 project\footnote{L1000 Connectivity Map perturbational profiles from Broad Institute LINCS Center for Transcriptomics: \texttt{https://www.ncbi.nlm.nih.gov/geo/query/acc.cgi?acc=GSE70138}}, \emph{without requiring access to the classification rules themselves}. Instead, the model learns a rich and powerful representation from the chemicals' effects on human cells and their unbiased, raw structure encoded in their \emph{molecular fingerprint}~\cite{cereto2015molecular}. Consequently, any chemical can be placed in one of the Verhaar classes if we know its effect on human cells and fingerprint (can be easily extracted using open-source tools and databases).

\section{Related work}

Existing research using the LINCS transcriptomic data is based in areas which include repurposing drugs, investigating their properties and predicting their adverse effects. Aliper et al.~\cite{aliper2016deep} used deep neural networks (DNNs) and support vector machines (SVMs) to learn from LINCS data corresponding to 678 drugs across three cell lines. They predicted 12 therapeutic use categories derived from the MeSH database and proposed a DNN confusion matrix approach for drug repurposing. Iwata et al.~\cite{iwata2017elucidating} also used the LINCS data to elucidate MOAs of bioactive compounds. They first performed pathway enrichment analysis to reveal similarly classified drugs, using previously adopted procedures (computing the $p$-value of two sets of genes intersecting), then predicted the target protein using known compound-target interactions as ground truth (via cell-based similarity search using the Pearson correlation coefficient) and finally revealed new such interactions.

Our approach is mostly similar to the one adopted by Wang et al.~\cite{wang2016drug}, who integrated gene expression (GE) data with cell multiplex-cytological (MC) profiles and the chemical structure (CS) of drugs in the form of fingerprints to predict adverse drug reactions (ADRs). The three data types were combined for feature selection, which yielded the 50 most predictive features that were used to learn an L1 regularized logistic regression model. The function coefficients were averaged across 200 runs to indicate feature importances. Extra Trees (ETs) classifiers were then trained for each ADR on GE, MC and CS data, using the 251 drugs shared among these three data sources. The authors discovered that the GE data had the highest prediction ability, whereas the CS and MC profiles showed similar AUROC (area under ROC) values. Additional findings showed that integrating MC profiles with other attributes did not significantly improve the classification performance, whereas combining CS features with GE data resulted in better predictions of ADRs. Unlike Wang et al., we do not restrict the model to a certain number of features and allow it to learn the best ones, providing only the raw data (entire gene expression vectors and fingerprints) to the classifier.

\section{Model and inputs}\label{s2}

\subsection{Dataset}

\begin{figure}[t]
    \centering
    \includegraphics[width=0.8\linewidth]{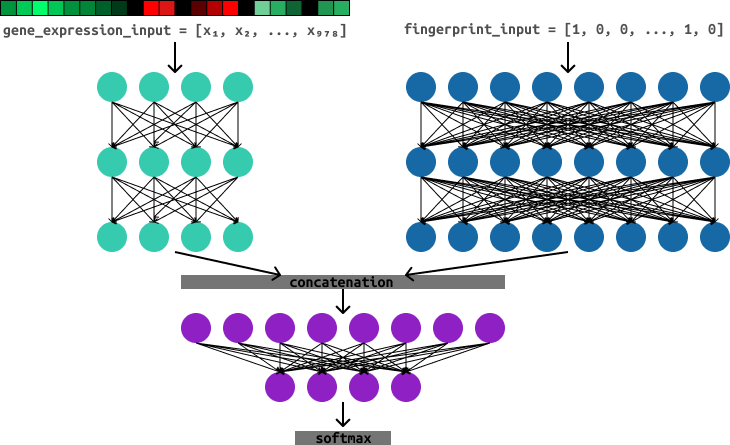}
    \caption{Illustration of the integrative architecture used to classify chemical compounds. Each example contains two inputs (vector of thresholded and averaged gene expression levels and fingerprint binary vector) and is placed into one of the four Verhaar classes.}
    \label{fig:concat}
\end{figure}

We used two types of data for classification: (a) \emph{gene expression levels} encoding the transcriptional responses of chemicals across human cell lines, represented by a vector of 978 landmark genes, and (b) the respective \emph{molecular fingerprints}, which are bit strings encoded by binary vectors of length 1024 (see Figure~\ref{fig:concat}). The goal is to place each compound (represented by the two modalities) in one of 4 Verhaar classes.

The gene expression levels were taken from the LINCS Phase II data. The entire dataset contains 118050 samples, corresponding to individual experiments with a single compound across 12328 genes, out of which 978 are \emph{landmark genes} (used to infer the remaining 11350). A portion of the dataset (53976 examples) has been labelled according to the Verhaar classification scheme~\cite{verhaar1992classifying}, which indicates the toxicity of compounds based on their modes of action. The labels correspond to \emph{four} Verhaar classes and a fifth category---chemicals that cannot be classified by the Verhaar rules.

The architecture was trained using the 14116 examples across the first four classes enumerated in Table~\ref{table:verhaar_distr}, each of them only containing perturbations corresponding to the 978 landmark genes.

\begin{table}[h]
\caption{Verhaar~\cite{verhaar1992classifying} class distribution for the labelled LINCS data, across the first four classes. The dataset is unbalanced and each chemical is used in multiple experiments.}
    \centering
\begin{tabular}{ c c c }
 \toprule
 Description & Experiments & Unique chemicals \\
 \midrule
Inert chemicals & 1092 & 26 \\
Less inert chemicals & 1949 & 37 \\
Reactive chemicals & 6474 & 104 \\
Specifically acting chemicals & 4601 & 31 \\
 \midrule
 \bf Total & \bf 14116 & \bf 198 \\
 \bottomrule
\label{table:verhaar_distr}
\end{tabular}
\end{table}

\subsection{Model architecture}

We use a multimodal neural network architecture to classify the chemicals by the Verhaar scheme (see Figure~\ref{fig:concat} for a graphical description). Two learning streams extract features separately from each of the modalities (gene expression levels and molecular fingerprint); the resulting 1D vectors are subsequently concatenated and a final, joint representation is learned at the tail of the model. The classification is achieved by a softmax layer.

\subsection{Data pre-processing}\label{datapp}

One chemical is always used in multiple experiments (ranging from 6 to 2310, most often 42)---and therefore examples---across the LINCS dataset. As a consequence, the unique chemicals represented in the labelled portion are far fewer than the number of experiments; the distribution across the four classes is shown in Table~\ref{table:verhaar_distr}. This implies that all training examples involving the same chemical would contain the same fingerprint data, which might result in \emph{overfitting} on the corresponding learning stream in the network. In order to avoid this issue, we \emph{summarized} the gene expression vectors from \emph{all experiments} that were carried out using the same chemical. The resulting summary is then used as part of one example, along with the corresponding fingerprint vector---this reduces the dataset size to \emph{198 examples}.

\begin{figure}[t]
    \centering
    \includegraphics[width=\linewidth]{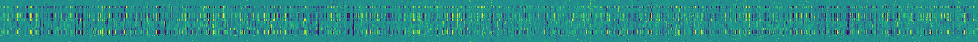} \\
    ${\big\rvert}$ \\
    $\textit{filter}\text{(standard deviation} \geq t \text{)}$ \\
    ${\Big\downarrow}$ \\
    \includegraphics[width=\linewidth]{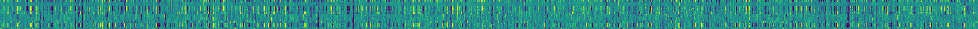} \\
    ${\big\rvert}$ \\
    $\textit{concat}\text{(element-wise avg, element-wise max)}$ \\
    ${\Big\downarrow}$ \\
    $[a_1, a_2, ..., a_{978}, m_{1}, m_{2}, ..., m_{978}]$ \\
    \caption{Schematic description of the summarization pipeline described in Section~\ref{datapp}. The input is the entire set of gene expression levels corresponding to experiments with a single chemical across the LINCS dataset. The pipeline outputs a single vector which contains the element-wise average and maximum values of the filtered gene expression vectors.}
    \label{fig:filtering}
\end{figure}

\paragraph{Summarization} Upon inspection of all experiments that use a single chemical, we noticed that some of the transcriptional response vectors do not exhibit high variance and are thus less informative than others, potentially even confounding the discrimination across classes. Each set of experiments was filtered by comparing the standard deviation of a gene expression level vector with a fixed threshold $t$---we found that $t = 1.25$ works best for this dataset. Finally, the summary is represented by a vector of length $2 \times \text{\emph{NUM\_LANDMARK\_GENES}}$ which represents the concatenation of the element-wise \emph{sum} and \emph{maximum} across the filtered experiments. A graphical description of the summarization pipeline is given in Figure~\ref{fig:filtering}.

\section{Experiments}

\subsection{Evaluation procedure}

We used $k$-fold cross-validation with $k = 5$ to assess the performance of our models. Both unimodal architectures were trained for 1000 epochs and the integrative model---for 2000 epochs. All models were trained on a batch size of 26, using the RMSprop optimizer~\cite{hinton2012neural}, with a learning rate of 0.005 and default hyperparameters~\footnote{\texttt{https://pytorch.org/docs/stable/optim.html\#torch.optim.RMSprop}} otherwise.

The class distribution across the LINCS labelled data is \emph{unbalanced} (see Table~\ref{table:verhaar_distr}), which makes accuracy an unsuitable metric---we therefore used a weighted version of the F1 score to evaluate model performance. Accordingly, all models were trained with a weighted cross-entropy loss function, where weights were inversely proportional to the relative class proportions.

\subsection{Model parameters}

Our unimodal baselines are both 3-layer perceptrons. Each layer has 256 hidden units and is followed by an ELU activation~\cite{he2015delving}, a batch normalization operation~\cite{ioffe2015batch} and dropout regularization~\cite{srivastava2014dropout} with $p = 0.25$ or $p = 0.5$ for the final layer.

The streams of the integrative model have the same layout as the baselines described above (256-D layers for fingerprints, 64-D for gene expression levels---see Section~\ref{s2} for details); the concatenation operation is regularized with dropout with $p = 0.25$. Finally, the tail of the model is formed of two fully-connected layers (256- and 32-D) with ELU activations and dropout layers with $p = 0.5$. Batch normalization is employed starting from the concatenation point.

\begin{table}
\caption{Results of 5-fold cross-validation for the unimodal and integrative architectures.}
    \centering
\begin{tabular}{ c c }
 \toprule
 Model & F1 score \\
 \midrule
 Majority class & 0.362 \\
 Gene expression & 0.606 \\
 Fingerprints & 0.803 \\
 \midrule
 \bf Fingerprints and gene expression & \bf 0.812 \\
 \bottomrule
\label{table:lincs_results}
\end{tabular}
\end{table}

\subsection{Results}

Table~\ref{table:lincs_results} shows that jointly learning from gene expression levels and molecular fingerprints results in a strong classification ability for the 4 Verhaar classes. This result improves over both unimodal architectures, suggesting that adding information about the effects of chemicals can further enhance the powerful discrimination capabilities of the fingerprint-only model, with respect to the Verhaar classification scheme.

This result is particularly important for classifying the unlabelled chemicals in LINCS. As a vital goal of this research direction, assessing the toxicity of \emph{any chemical} could become a simple process by using the integrative model. The next essential step is to validate the predictions on the unlabelled chemicals experimentally and using expert knowledge.

\bibliographystyle{plain}
\bibliography{nips_2018}

\appendix

\end{document}